\documentclass{article}
\usepackage{spconf,amsmath,graphicx}

\usepackage{amssymb,amsfonts,mathtools}
\usepackage{xcolor}
\usepackage{tikz}
\usepackage{pgfplots}
\DeclareUnicodeCharacter{2212}{−}
\usepgfplotslibrary{groupplots,dateplot}
\usetikzlibrary{patterns,shapes.arrows}
\pgfplotsset{compat=newest}
\usepackage{epsfig}
\usepackage{svg}
\usepackage[linesnumbered,ruled,vlined]{algorithm2e}

\usepackage{algorithmicx}
\usepackage{algpseudocode}
\usepackage[T1]{fontenc}
\usepackage{textcomp}
\usepackage{upquote}
\usepackage{multirow}
\usepackage{booktabs}
\usepackage{tablefootnote}
\usepackage{caption}
\usepackage{subfigure}
\usepackage{tabularx}
\usepackage{makecell}
\usepackage{adjustbox}
\usepackage{float}
\usepackage{framed}
\usepackage{ragged2e}
\usepackage{balance}
\usepackage{fancyhdr}
\usepackage{listings}
\usepackage{comment}
\usepackage{url}
\usepackage{lastpage}
\usepackage{academicons}
\usepackage{bm}

\usepackage{hyperref}
\usepackage{cite}

\DeclareRobustCommand{\circlednum}[1]{%
    \tikz[baseline=(char.base)]{
        \node[
            shape=circle,
            fill=black,
            text=white,
            inner sep=1pt
        ] (char) {\fontsize{7pt}{7pt}\selectfont #1};
    }%
}
\newcommand{\supc}[1]{%
    (\tikz[baseline=(char.base)]{
        \node[shape=circle, fill=black, text=white, inner sep=0.5pt] (char) {\small #1};
    })
}

\newcommand{\papername}{FRESHLATENT}
\title{\papername{}: Channel-Aware Latent Adaptation for Resource-Constrained Embodied VLM Perception}

\name{%
\begin{tabular}{c}
Rajat Bhattacharjya\textsuperscript{1},\;
Minwoo Kim\textsuperscript{2},\;
Arnab Sarkar\textsuperscript{3},\;
Tamoghno Das\textsuperscript{1},
\\
Sing-Yao Wu\textsuperscript{1},\;
Eli Bozorgzadeh\textsuperscript{1},\;
Marco Levorato\textsuperscript{1},\;
Nikil Dutt\textsuperscript{1}
\end{tabular}%
\thanks{This work has been submitted to the IEEE for possible publication.
Copyright may be transferred without notice, after which this version may
no longer be accessible. Authors' version posted for personal use and not
for redistribution.}%
}

\address{%
\textsuperscript{1}University of California, Irvine, USA; \
\textsuperscript{2}Kookmin University, Seoul, South Korea\\
\textsuperscript{3}Indian Institute of Technology, Kharagpur, India\\
\texttt{Corresponding author: rajatb1@uci.edu}}

\begin{document}
\ninept

\maketitle

\begin{abstract}
Mission-critical UAVs increasingly rely on split vision-language model (VLM) perception under tight onboard-resource and wireless-communication constraints. 
However, corruption of transmitted intermediate features creates a deployment mismatch for clean-trained split interfaces, while stronger channel-aware codecs can impose substantial onboard cost. 
We present \texttt{FreshLatent}, a lightweight channel-aware latent adapter that trains a power-normalized encoder--decoder through wireless corruption while keeping the surrounding VLM frozen. 
We formulate deployment around a mission-conditioned perception requirement and embedded interface cost, linking channel quality and communication budget to the operating conditions under which perception remains usable. 
At 0\,dB and the tightest communication budget, \texttt{FreshLatent} improves gIoU and cIoU over clean split compression by 20.79 and 20.87 points, respectively. 
At the most adverse evaluated SNR (0 dB), across all three communication budgets, \texttt{FreshLatent} recovers 63.5–69.1\% of the gIoU improvement achieved by a much heavier, range-trained feature-JSCC codec.
On an NVIDIA Jetson AGX Xavier in 10-W mode, \texttt{FreshLatent} uses 37--40$\times$ fewer encoder parameters, 7.7--9.9$\times$ lower edge-interface latency, and 8.8--10.0$\times$ lower edge-interface energy than the heavier codec.
Together, these results show that lightweight channel-aware adaptation can recover a substantial fraction of the robustness of a much larger communication interface while broadening quality-valid operation under constrained wireless conditions.
\end{abstract}

\vspace{-2ex}
\section{Introduction}
\begin{figure}[t]
    \centering

    \includegraphics[width=0.9\columnwidth]{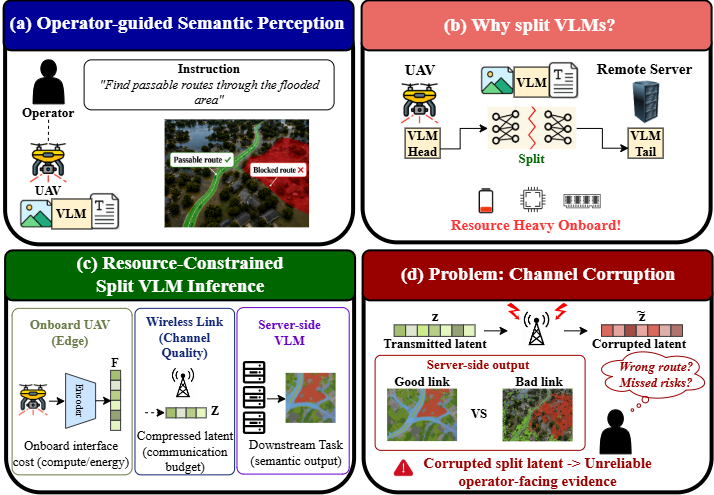}
    \captionsetup{
        font=footnotesize,
        labelfont={bf,footnotesize}
    }
    \caption{Motivation for \texttt{FreshLatent}. 
    (a) Mission-driven UAVs rely on query-conditioned semantic perception.
    (b) Onboard resource constraints motivate UAV--server split VLM execution.
    (c) Split inference couples onboard interface cost, communication budget, and wireless channel quality.
    (d) Channel corruption can degrade downstream perception and render the returned evidence insufficient for the required task quality, motivating channel-aware latent adaptation.}
    \label{fig:motivation}
    \vspace{-5ex}
\end{figure}

\vspace{-2ex}
Mission-critical UAVs, such as those used in disaster response, are increasingly moving from passive image collection toward operator-guided semantic perception~\cite{bhattacharjya2025avery,cladera2025air,yaq,ke2026pixdlm, bhattacharjya2026floodreasonbench}.
An operator may request \textit{``Find passable routes through the flooded area''} or \textit{``Segment damaged regions near the collapsed structure,''} using the returned spatial evidence to guide mission-level decisions.
As illustrated in Fig.~\ref{fig:motivation}(a), such systems therefore require query-conditioned perception that remains sufficiently reliable for human-supervised operation.

Vision-language models (VLMs) provide a natural interface for such perception, but their computation and memory requirements challenge fully onboard deployment~\cite{vlmc,li2025benchmark}.
Split computing~\cite{matsubara2022split, bakhtiarnia2023dynamic} offers an alternative in which the UAV performs early visual processing and transmits an intermediate representation to a remote server for downstream reasoning and segmentation (Fig.~\ref{fig:motivation}(b)--(c)).
Unlike conventional image offloading, the wireless link now carries an \emph{internal model representation} that must remain compatible with the downstream VLM.
Reducing its dimensionality lowers communication and interface cost, but channel corruption before reconstruction can propagate through the frozen downstream model and degrade the returned segmentation (Fig.~\ref{fig:motivation}(d)).

Existing split-computing bottlenecks primarily optimize intermediate representations for clean transfer~\cite{bottlefit,bhattacharjya2025avery}, while channel-aware representation learning and joint source--channel coding (JSCC)~\cite{bourtsoulatze2019deepjscc, gunduz2024joint} improve robustness to wireless distortion.
Recent edge--cloud VLM systems further compress or progressively
refine intermediate visual representations under communication
constraints~\cite{hsu2026progressive,heidari2026compressing}.
However, a resource-constrained split VLM raises a complementary question:
\emph{how can an existing lightweight model interface be made robust to wireless corruption without substantially increasing edge overhead?}
A more expressive channel-aware codec may improve robustness, but then gains may also reflect additional codec capacity and computation.

We present \texttt{FreshLatent}, a lightweight channel-aware
adaptation of the split-VLM interface. Rather than increasing codec
capacity, \texttt{FreshLatent} trains a power-normalized latent
encoder--decoder through wireless corruption to reconstruct the
pretrained intermediate representation while leaving the surrounding
VLM unchanged. This lets us separate two sources of robustness:
adaptation of an existing lightweight interface to the channel and
additional capacity from a heavier feature-JSCC codec. We further
associate downstream perception with a mission-dependent minimum
quality requirement, allowing the measured channel--communication
tradeoff to be interpreted together with onboard codec cost.




Our contributions are threefold: \textbf{(i)} we formulate wireless split-VLM
operation through a mission-conditioned perception requirement and
embedded resource feasibility; \textbf{(ii)} we develop \texttt{FreshLatent},
a lightweight channel-aware adaptation of a frozen VLM interface; and \textbf{(iii)} we show that, under matched range training, a lightweight channel-aware interface recovers a substantial fraction of the poor-channel robustness gain of a much larger feature-JSCC codec while requiring substantially lower embedded interface cost.

\vspace{-3mm}
\section{System Model and Problem Formulation}
\label{sec:problem}

\begin{figure}[t]
    \centering
    \includegraphics[width=0.9\columnwidth]{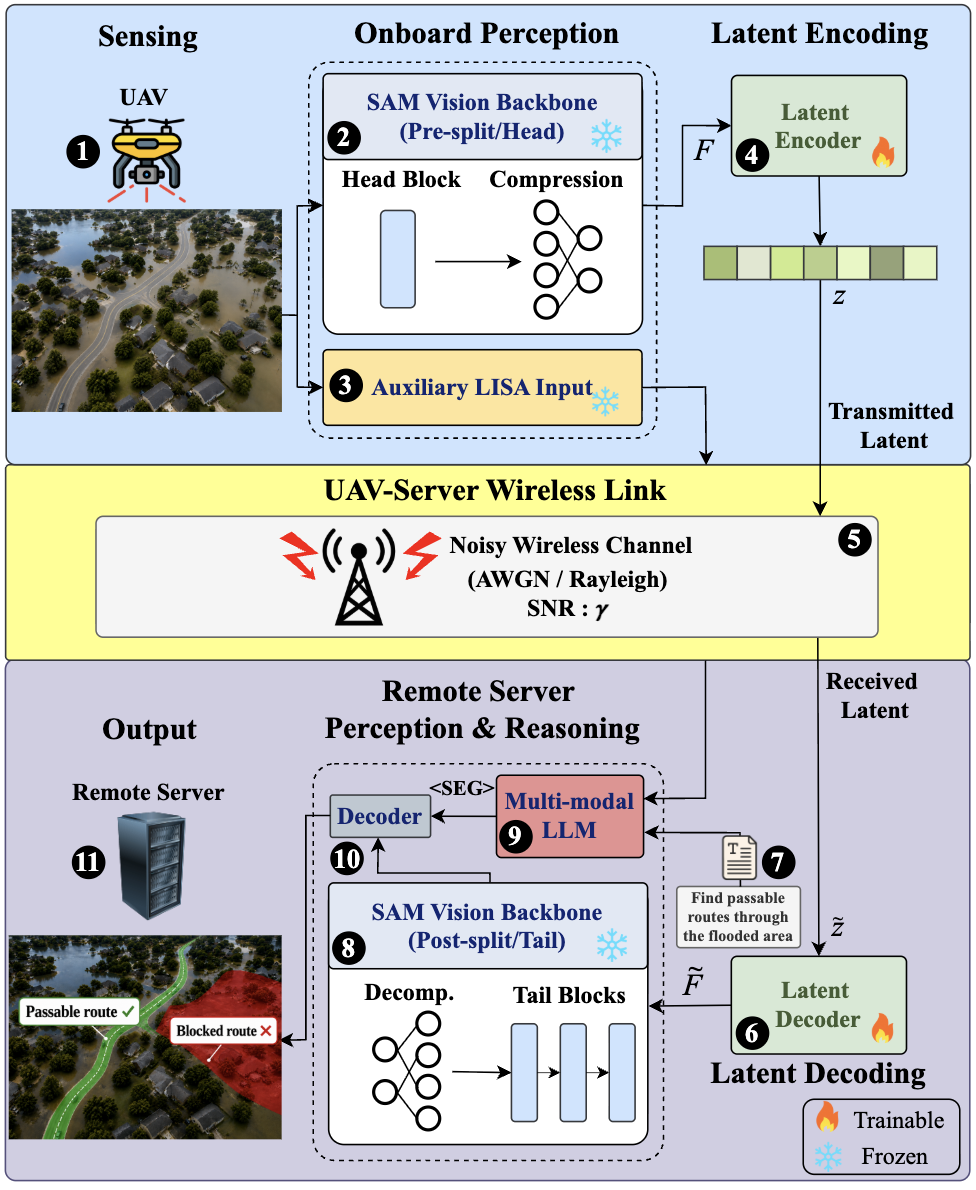}
    \vspace{-2ex}
    \captionsetup{
        font=footnotesize,
        labelfont={bf,footnotesize}
    }
    \caption{Overview of \texttt{FreshLatent} in the split LISA pipeline.
    The UAV extracts an intermediate SAM feature $F$ and transmits a
    compressed latent $z$ over the wireless link. The server reconstructs
    $\tilde{F}$, which is consumed together with the operator query by the
    frozen downstream LISA pipeline. \texttt{FreshLatent} adapts only the
    communication interface spanning \circlednum{4}--\circlednum{6}; the surrounding perception pipeline remains unchanged.}
    \label{fig:freshlatent_system}
    \vspace{-6mm}
\end{figure}

\vspace{-3mm}
We consider the human-supervised split-VLM setting in
Fig.~\ref{fig:freshlatent_system}, where a resource-constrained UAV
provides query-conditioned spatial evidence to a remote operator.
The UAV performs early visual processing and transmits an intermediate
representation over a wireless link, while a remote server completes
reasoning and segmentation.
Since this intermediate representation is consumed internally by the downstream
VLM, channel quality, communication demand, and onboard resource cost
jointly shape whether an operating point is suitable for the mission.

\vspace{-4mm}
\subsection{Operator-Guided Split VLM Perception}

Let $x$ denote the UAV image~\supc{1} and $q$ a mission-specific
operator query~\supc{7}.
The onboard visual prefix $E_{\mathrm{pre}}$~\supc{2} produces the
split feature
\begin{equation}
    F = E_{\mathrm{pre}}(x),
    \qquad
    F \in \mathbb{R}^{H\times W\times d},
    \label{eq:split_feature}
\end{equation}
where $H$, $W$, and $d$ denote its spatial and channel dimensions.

At latent rate $r$, a rate-specific encoder~\supc{4} maps $F$ to
\begin{equation}
    z=f_{\theta_r}(F),
    \qquad
    z\in\mathbb{R}^{H\times W\times d_z},
    \qquad
    d_z=\lfloor rd\rfloor ,
    \label{eq:latent}
\end{equation}
where a separate encoder--decoder pair is used for each supported
latent rate.
The latent is power-normalized and transmitted through the wireless
link~\supc{5}, after which the server-side decoder~\supc{6}
reconstructs
\begin{equation}
    \tilde{F}
    =
    g_{\phi_r}
    \left(
        \mathcal{H}
        \left(
            \mathcal{P}(z);
            \gamma,\omega
        \right)
    \right),
    \label{eq:wireless_interface}
\end{equation}
where $\gamma$ denotes channel SNR and $\omega$ the stochastic channel
realization.

The reconstructed feature, operator query, and unchanged auxiliary
representation $c$~\supc{3} are consumed by the frozen server-side
VLM~\supc{8}--\supc{10} as
$\hat{m}=G_{\mathrm{server}}(\tilde{F},c,q)$, where $\hat{m}$ is the
query-conditioned segmentation presented to the operator~\supc{11}.

Unlike image offloading, the wireless link therefore carries an
\emph{internal model representation}: distortion of $z$ can alter
$\tilde{F}$ and propagate through the frozen downstream computation.
For the spatially preserving latent used by \texttt{FreshLatent}, each
real-valued latent element occupies one real channel use, giving
\begin{equation}
    K(r)=HWd_z=HW\lfloor rd\rfloor .
    \label{eq:symbol_budget}
\end{equation}
We use $K$ as the common communication-budget variable when comparing
different transmission architectures; methods operating on images or
other representations are matched by channel uses rather than by the
feature-specific rate $r$.

\vspace{-3mm}
\subsection{Mission-Conditioned Task Validity}

Let $Q^{M}(K,\gamma)$ denote the downstream perception quality of
method $M$ under communication budget $K$ and channel SNR $\gamma$.
Since we do not directly measure operator decisions or mission success,
we model mission context $\mu$ through a semantic utility $U_{\mu}(Q)$,
instantiated as a minimum perception requirement:
\begin{equation}
    U_{\mu}(Q)
    =
    \mathbf{1}\!\left[Q\geq\tau_{\mu}\right],
    \label{eq:semantic_utility}
\end{equation}
where $\tau_{\mu}$ denotes the minimum acceptable perception quality.
Rather than assigning a universal value to $\tau_{\mu}$, we evaluate
the quality requirement across a range of representative thresholds.

Resource feasibility must also hold.
Let $\mathbf{C}^{M}(K)$ denote the onboard interface cost of method
$M$ at communication budget $K$, and let $\mathbf{B}_{\mu}$ denote
the corresponding deployment resource budget.
The cost vector may include interface parameter footprint, codec
latency, and codec energy, with $\preceq$ denoting component-wise
satisfaction of these resource limits.
We define a task-valid operating point as
\begin{equation}
    V^{M}(K,\gamma;\mu)
    =
    U_{\mu}\!\left(Q^{M}(K,\gamma)\right)
    \,
    \mathbf{1}
    \left[
        \mathbf{C}^{M}(K)\preceq\mathbf{B}_{\mu}
    \right].
    \label{eq:task_validity}
\end{equation}

This formulation separates the two conditions required for deployment:
sufficient perception quality and feasible onboard interface cost.
In Sec.~\ref{sec:operating_space}, we evaluate the quality component
across representative $\tau_{\mu}$, while
Sec.~\ref{sec:embedded_results} measures
$\mathbf{C}^{M}(K)$.
A specific deployment can combine these measurements with its own
resource budget $\mathbf{B}_{\mu}$ to determine task-valid operating
points.

\vspace{-1.5mm}

\vspace{-2ex}
\section{\texttt{FreshLatent}: Channel-Aware Latent Adaptation}
\label{sec:method}
\vspace{-3ex}
\texttt{FreshLatent} adapts the communication interface
\supc{4}--\supc{6} in Fig.~\ref{fig:freshlatent_system}, while the
surrounding SAM and LISA components remain frozen.
The interface compresses the intermediate visual representation,
transports it through the wireless channel, and reconstructs the
feature expected by the unchanged downstream VLM.

\vspace{-3mm}
\subsection{Lightweight Latent Interface}
\vspace{-1mm}
We split the SAM~\cite{sam} ViT-H image encoder after its first
transformer block, following~\cite{bhattacharjya2025avery}.
For each supported latent rate $r$, a separate lightweight bottleneck
compresses
$F\in\mathbb{R}^{H\times W\times d}$ to
$d_z=\lfloor rd\rfloor$ channels.
The encoder and decoder apply shared linear projections independently at each spatial location,
$z_{i,j}=W_{e,r}F_{i,j}$ and
$\tilde{F}_{i,j}=W_{d,r}\tilde{z}_{i,j}$, where
$W_{e,r}\in\mathbb{R}^{d_z\times d}$ and
$W_{d,r}\in\mathbb{R}^{d\times d_z}$.
Because the projection acts only along the channel dimension, the
$H\times W$ spatial organization of the pretrained SAM feature is
preserved while reducing the transmitted representation size.

Before transmission, the latent is normalized to satisfy a unit
average-power constraint:

\begin{equation}
    \bar{z}
    =
    \mathcal{P}(z)
    =
    \sqrt{K(r)}
    \frac{z}{\|z\|_2},
    \qquad
    \frac{1}{K(r)}
    \|\bar{z}\|_2^2=1,
    \label{eq:power_norm}
\end{equation}
where $K(r)$ is defined in Eq.~\eqref{eq:symbol_budget}.
The received latent is $\tilde{z}=\mathcal{H}(\bar{z};\gamma,\omega)$; for AWGN,
$\tilde{z}=\bar{z}+n$, where
$n\sim\mathcal{N}\!\left(0,10^{-\gamma_{\mathrm{dB}}/10}I\right)$.

The server-side linear decoder then maps $\tilde z$ back to the
feature dimensionality expected by the frozen post-split computation. Each real latent element is counted as one channel use.

\vspace{-3mm}
\subsection{Channel-Aware Reconstruction Learning}
\vspace{-1mm}
A clean-trained split bottleneck learns to reconstruct $F$ without
exposing its latent representation to wireless corruption.
At deployment, however, the decoder operates on the channel-corrupted
latent $\tilde z$.
\texttt{FreshLatent} addresses this mismatch by placing the wireless
channel directly inside the reconstruction path during training.

For each rate $r$, the latent encoder and decoder are optimized as
\begin{equation}
\begin{split}
(\theta_r^*,\phi_r^*)
=
\arg\min_{\theta_r,\phi_r}
\;
\mathbb{E}_{F,\gamma_{\mathrm{tr}},\omega}
\left[
\frac{1}{HWd}
\left\|
\tilde{F}-F
\right\|_F^2
\right],
\end{split}
\label{eq:freshlatent_objective}
\end{equation}
where $\tilde F$ is obtained through the channel-aware path above,
$\gamma_{\mathrm{tr}}$ denotes the training SNR, and $\omega$ denotes
the sampled channel realization.
Only the latent encoder and decoder are optimized; the feature
extractor and downstream VLM remain frozen.

The reconstruction target is the original pretrained feature $F$.
Thus, rather than changing the downstream reasoning model,
\texttt{FreshLatent} learns a channel-aware interface whose
reconstructed output remains compatible with the internal
representation expected by the frozen VLM.

The distribution of $\gamma_{\mathrm{tr}}$ determines the channel
conditions observed during training.
For \texttt{FL-Fixed}, training is concentrated at 13\,dB.
The primary \texttt{FreshLatent} model instead samples
$\gamma_{\mathrm{tr}}$ across the predefined training range.
For each latent rate, one range-trained \texttt{FreshLatent} checkpoint
is evaluated across all channel conditions.
For comparison, \texttt{SNR-Specialist} is formed from independently
trained fixed-SNR checkpoints, selecting the best available specialist
at each evaluation SNR.
It therefore measures how closely range training approaches
condition-specific specialization at each latent rate.
\vspace{-4mm}
\section{Experiments and Evaluation}
\label{sec:experiments}

\vspace{-2mm}
\subsection{Experimental Setup}
\vspace{-1mm}
\begin{figure*}[t]
\centering
\definecolor{cBlue}{RGB}{31,119,180}
\definecolor{cOrange}{RGB}{255,127,14}
\definecolor{cGreen}{RGB}{44,160,44}
\definecolor{cRed}{RGB}{214,39,40}
\definecolor{cPurple}{RGB}{148,103,189}
\definecolor{cBrown}{RGB}{140,86,75}
\definecolor{cPink}{RGB}{227,119,194}
\definecolor{cCyan}{RGB}{23,190,207}
\begin{tikzpicture}
\begin{groupplot}[
  group style={group size=3 by 2, horizontal sep=0.3cm, vertical sep=1.7cm,
               yticklabels at=edge left, ylabels at=edge left},
  width=0.34\textwidth, height=4.1cm,
  xmin=-0.8, xmax=25.8, xtick={0,6,12,18,25},
  xlabel={SNR (dB)},
  grid=major, grid style={gray!20, line width=0.3pt},
  tick align=inside, tick label style={font=\scriptsize},
  label style={font=\small}, title style={font=\small, yshift=-1ex},
  every axis plot/.append style={line width=0.9pt},
  legend style={font=\footnotesize, draw=none, /tikz/every even column/.append style={column sep=0.22cm}},
]
\nextgroupplot[title={(a) $K{=}0.262$M}, ymin=0.26, ymax=0.81, ytick={0.3,0.4,0.5,0.6,0.7,0.8}, ylabel={gIoU}]
\addplot[cBlue, dashed, mark=*, mark size=1.5pt, mark options={solid}] coordinates {(0,0.2816) (3,0.4209) (6,0.5380) (9,0.6274) (12,0.6695) (15,0.6962) (18,0.7184) (21,0.7289) (25,0.7429)};
\addplot[cOrange, solid, mark=square*, mark size=1.4pt, mark options={solid}] coordinates {(0,0.4650) (3,0.5810) (6,0.6388) (9,0.6728) (12,0.6916) (15,0.7104) (18,0.7220) (21,0.7338) (25,0.7460)};
\addplot[cGreen, solid, mark=triangle*, mark size=1.8pt, mark options={solid}] coordinates {(0,0.4895) (3,0.5834) (6,0.6281) (9,0.6707) (12,0.6846) (15,0.7070) (18,0.7240) (21,0.7359) (25,0.7461)};
\addplot[cRed, solid, mark=diamond*, mark size=1.9pt, mark options={solid}] coordinates {(0,0.4988) (3,0.5833) (6,0.6388) (9,0.6728) (12,0.6916) (15,0.7104) (18,0.7235) (21,0.7368) (25,0.7460)};
\addplot[cPurple, dotted, mark=pentagon*, mark size=1.2pt, mark options={solid}] coordinates {(0,0.5747) (3,0.6448) (6,0.6671) (9,0.6981) (12,0.7143) (15,0.7219) (18,0.7309) (21,0.7273) (25,0.7294)};
\addplot[cCyan, dotted, mark=square*, mark size=1.3pt, mark options={solid}] coordinates {(0,0.6090) (3,0.6448) (6,0.6734) (9,0.6966) (12,0.7081) (15,0.7147) (18,0.7184) (21,0.7215) (25,0.7209)};
\addplot[cBrown, dashed, mark=pentagon*, mark size=1.7pt, mark options={solid}] coordinates {(0,0.5071) (3,0.5734) (6,0.6152) (9,0.6472) (12,0.6897) (15,0.7106) (18,0.7252) (21,0.7340) (25,0.7342)};
\addplot[cPink, dashdotted, mark=x, mark size=2.2pt, mark options={solid}] coordinates {(0,0.0595) (3,0.0595) (6,0.0595) (9,0.0595) (12,0.0595) (15,0.7882) (18,0.7882) (21,0.7882) (25,0.7882)};
\nextgroupplot[title={(b) $K{=}0.524$M}, ymin=0.26, ymax=0.81, ytick={0.3,0.4,0.5,0.6,0.7,0.8}]
\addplot[cBlue, dashed, mark=*, mark size=1.5pt, mark options={solid}] coordinates {(0,0.4622) (3,0.5825) (6,0.6521) (9,0.6802) (12,0.7021) (15,0.7165) (18,0.7350) (21,0.7486) (25,0.7635)};
\addplot[cOrange, solid, mark=square*, mark size=1.4pt, mark options={solid}] coordinates {(0,0.5617) (3,0.6482) (6,0.6742) (9,0.7016) (12,0.7142) (15,0.7337) (18,0.7470) (21,0.7596) (25,0.7648)};
\addplot[cGreen, solid, mark=triangle*, mark size=1.8pt, mark options={solid}] coordinates {(0,0.5917) (3,0.6494) (6,0.6734) (9,0.7010) (12,0.7196) (15,0.7355) (18,0.7513) (21,0.7631) (25,0.7628)};
\addplot[cRed, solid, mark=diamond*, mark size=1.9pt, mark options={solid}] coordinates {(0,0.6013) (3,0.6498) (6,0.6759) (9,0.7016) (12,0.7213) (15,0.7384) (18,0.7533) (21,0.7612) (25,0.7676)};
\addplot[cPurple, dotted, mark=pentagon*, mark size=1.2pt, mark options={solid}] coordinates {(0,0.6394) (3,0.6681) (6,0.6951) (9,0.7317) (12,0.7360) (15,0.7408) (18,0.7417) (21,0.7441) (25,0.7439)};
\addplot[cCyan, dotted, mark=square*, mark size=1.3pt, mark options={solid}] coordinates {(0,0.6495) (3,0.6782) (6,0.7043) (9,0.7228) (12,0.7249) (15,0.7364) (18,0.7377) (21,0.7332) (25,0.7341)};
\addplot[cBrown, dashed, mark=pentagon*, mark size=1.7pt, mark options={solid}] coordinates {(0,0.5564) (3,0.5845) (6,0.6309) (9,0.6648) (12,0.6967) (15,0.7181) (18,0.7326) (21,0.7359) (25,0.7377)};
\addplot[cPink, dashdotted, mark=x, mark size=2.2pt, mark options={solid}] coordinates {(0,0.0595) (3,0.0595) (6,0.0595) (9,0.0595) (12,0.0595) (15,0.7895) (18,0.7895) (21,0.7895) (25,0.7895)};
\nextgroupplot[title={(c) $K{=}1.311$M}, ymin=0.26, ymax=0.81, ytick={0.3,0.4,0.5,0.6,0.7,0.8}]
\addplot[cBlue, dashed, mark=*, mark size=1.5pt, mark options={solid}] coordinates {(0,0.5861) (3,0.6714) (6,0.7096) (9,0.7257) (12,0.7421) (15,0.7513) (18,0.7628) (21,0.7775) (25,0.7894)};
\addplot[cOrange, solid, mark=square*, mark size=1.4pt, mark options={solid}] coordinates {(0,0.6495) (3,0.6832) (6,0.7099) (9,0.7263) (12,0.7407) (15,0.7539) (18,0.7700) (21,0.7763) (25,0.7843)};
\addplot[cGreen, solid, mark=triangle*, mark size=1.8pt, mark options={solid}] coordinates {(0,0.6666) (3,0.6918) (6,0.7091) (9,0.7253) (12,0.7453) (15,0.7547) (18,0.7718) (21,0.7737) (25,0.7742)};
\addplot[cRed, solid, mark=diamond*, mark size=1.9pt, mark options={solid}] coordinates {(0,0.6628) (3,0.6895) (6,0.7118) (9,0.7282) (12,0.7436) (15,0.7574) (18,0.7700) (21,0.7791) (25,0.7879)};
\addplot[cPurple, dotted, mark=pentagon*, mark size=1.2pt, mark options={solid}] coordinates {(0,0.6938) (3,0.7250) (6,0.7484) (9,0.7709) (12,0.7819) (15,0.7848) (18,0.7835) (21,0.7866) (25,0.7869)};
\addplot[cCyan, dotted, mark=square*, mark size=1.3pt, mark options={solid}] coordinates {(0,0.7096) (3,0.7339) (6,0.7524) (9,0.7684) (12,0.7758) (15,0.7772) (18,0.7779) (21,0.7785) (25,0.7786)};
\addplot[cBrown, dashed, mark=pentagon*, mark size=1.7pt, mark options={solid}] coordinates {(0,0.5767) (3,0.6183) (6,0.6475) (9,0.6695) (12,0.6948) (15,0.7201) (18,0.7320) (21,0.7361) (25,0.7410)};
\addplot[cPink, dashdotted, mark=x, mark size=2.2pt, mark options={solid}] coordinates {(0,0.0595) (3,0.0595) (6,0.0595) (9,0.0595) (12,0.0595) (15,0.7882) (18,0.7882) (21,0.7882) (25,0.7882)};
\nextgroupplot[title={(d) $K{=}0.262$M}, ymin=0.36, ymax=0.91, ytick={0.4,0.5,0.6,0.7,0.8,0.9}, ylabel={cIoU}]
\addplot[cBlue, dashed, mark=*, mark size=1.5pt, mark options={solid}] coordinates {(0,0.3842) (3,0.5292) (6,0.6375) (9,0.7288) (12,0.7739) (15,0.8109) (18,0.8340) (21,0.8458) (25,0.8589)};
\addplot[cOrange, solid, mark=square*, mark size=1.4pt, mark options={solid}] coordinates {(0,0.5608) (3,0.6663) (6,0.7472) (9,0.7846) (12,0.8075) (15,0.8334) (18,0.8436) (21,0.8509) (25,0.8616)};
\addplot[cGreen, solid, mark=triangle*, mark size=1.8pt, mark options={solid}] coordinates {(0,0.5929) (3,0.6825) (6,0.7500) (9,0.7751) (12,0.7948) (15,0.8221) (18,0.8396) (21,0.8491) (25,0.8548)};
\addplot[cRed, solid, mark=diamond*, mark size=1.9pt, mark options={solid}] coordinates {(0,0.6047) (3,0.6898) (6,0.7472) (9,0.7846) (12,0.8075) (15,0.8334) (18,0.8403) (21,0.8514) (25,0.8616)};
\addplot[cPurple, dotted, mark=pentagon*, mark size=1.2pt, mark options={solid}] coordinates {(0,0.6963) (3,0.7586) (6,0.7898) (9,0.8109) (12,0.8292) (15,0.8353) (18,0.8393) (21,0.8404) (25,0.8396)};
\addplot[cCyan, dotted, mark=square*, mark size=1.3pt, mark options={solid}] coordinates {(0,0.7230) (3,0.7731) (6,0.7914) (9,0.8199) (12,0.8258) (15,0.8294) (18,0.8304) (21,0.8308) (25,0.8304)};
\addplot[cBrown, dashed, mark=pentagon*, mark size=1.7pt, mark options={solid}] coordinates {(0,0.6230) (3,0.6638) (6,0.7268) (9,0.7431) (12,0.8015) (15,0.8300) (18,0.8455) (21,0.8530) (25,0.8551)};
\addplot[cPink, dashdotted, mark=x, mark size=2.2pt, mark options={solid}] coordinates {(0,0.1245) (3,0.1245) (6,0.1245) (9,0.1245) (12,0.1245) (15,0.8887) (18,0.8887) (21,0.8887) (25,0.8887)};
\nextgroupplot[title={(e) $K{=}0.524$M}, ymin=0.36, ymax=0.91, ytick={0.4,0.5,0.6,0.7,0.8,0.9}]
\addplot[cBlue, dashed, mark=*, mark size=1.5pt, mark options={solid}] coordinates {(0,0.5573) (3,0.6958) (6,0.7689) (9,0.8005) (12,0.8262) (15,0.8372) (18,0.8497) (21,0.8618) (25,0.8730)};
\addplot[cOrange, solid, mark=square*, mark size=1.4pt, mark options={solid}] coordinates {(0,0.6616) (3,0.7616) (6,0.7907) (9,0.8224) (12,0.8341) (15,0.8491) (18,0.8612) (21,0.8705) (25,0.8761)};
\addplot[cGreen, solid, mark=triangle*, mark size=1.8pt, mark options={solid}] coordinates {(0,0.6941) (3,0.7678) (6,0.7882) (9,0.8188) (12,0.8377) (15,0.8536) (18,0.8645) (21,0.8706) (25,0.8738)};
\addplot[cRed, solid, mark=diamond*, mark size=1.9pt, mark options={solid}] coordinates {(0,0.7071) (3,0.7614) (6,0.7958) (9,0.8224) (12,0.8426) (15,0.8568) (18,0.8677) (21,0.8729) (25,0.8760)};
\addplot[cPurple, dotted, mark=pentagon*, mark size=1.2pt, mark options={solid}] coordinates {(0,0.7548) (3,0.7888) (6,0.8095) (9,0.8414) (12,0.8520) (15,0.8533) (18,0.8558) (21,0.8569) (25,0.8567)};
\addplot[cCyan, dotted, mark=square*, mark size=1.3pt, mark options={solid}] coordinates {(0,0.7625) (3,0.7956) (6,0.8235) (9,0.8376) (12,0.8406) (15,0.8441) (18,0.8445) (21,0.8465) (25,0.8469)};
\addplot[cBrown, dashed, mark=pentagon*, mark size=1.7pt, mark options={solid}] coordinates {(0,0.6343) (3,0.7112) (6,0.7345) (9,0.7731) (12,0.8009) (15,0.8397) (18,0.8540) (21,0.8577) (25,0.8587)};
\addplot[cPink, dashdotted, mark=x, mark size=2.2pt, mark options={solid}] coordinates {(0,0.1245) (3,0.1245) (6,0.1245) (9,0.1245) (12,0.1245) (15,0.8951) (18,0.8951) (21,0.8951) (25,0.8951)};
\nextgroupplot[title={(f) $K{=}1.311$M}, ymin=0.36, ymax=0.91, ytick={0.4,0.5,0.6,0.7,0.8,0.9}, legend to name=fig4legend, legend columns=4]
\addplot[cBlue, dashed, mark=*, mark size=1.5pt, mark options={solid}] coordinates {(0,0.6899) (3,0.7864) (6,0.8274) (9,0.8503) (12,0.8613) (15,0.8680) (18,0.8743) (21,0.8817) (25,0.8869)};
\addlegendentry{Clean}
\addplot[cOrange, solid, mark=square*, mark size=1.4pt, mark options={solid}] coordinates {(0,0.7901) (3,0.8116) (6,0.8395) (9,0.8541) (12,0.8605) (15,0.8693) (18,0.8793) (21,0.8847) (25,0.8854)};
\addlegendentry{FL-Fixed}
\addplot[cGreen, solid, mark=triangle*, mark size=1.8pt, mark options={solid}] coordinates {(0,0.7755) (3,0.8114) (6,0.8340) (9,0.8441) (12,0.8642) (15,0.8707) (18,0.8796) (21,0.8816) (25,0.8823)};
\addlegendentry{FreshLatent}
\addplot[cRed, solid, mark=diamond*, mark size=1.9pt, mark options={solid}] coordinates {(0,0.7796) (3,0.8050) (6,0.8364) (9,0.8529) (12,0.8609) (15,0.8700) (18,0.8793) (21,0.8845) (25,0.8867)};
\addlegendentry{SNR-Specialist}
\addplot[cPurple, dotted, mark=pentagon*, mark size=1.2pt, mark options={solid}] coordinates {(0,0.8151) (3,0.8478) (6,0.8669) (9,0.8822) (12,0.8863) (15,0.8868) (18,0.8854) (21,0.8860) (25,0.8863)};
\addlegendentry{FFMNet-JSCC}
\addplot[cCyan, dotted, mark=square*, mark size=1.3pt, mark options={solid}] coordinates {(0,0.8362) (3,0.8602) (6,0.8713) (9,0.8830) (12,0.8840) (15,0.8845) (18,0.8847) (21,0.8848) (25,0.8848)};
\addlegendentry{FFMNet-JSCC-Range}
\addplot[cBrown, dashed, mark=pentagon*, mark size=1.7pt, mark options={solid}] coordinates {(0,0.6717) (3,0.7144) (6,0.7609) (9,0.7724) (12,0.8051) (15,0.8301) (18,0.8455) (21,0.8552) (25,0.8611)};
\addlegendentry{DeepJSCC-Image}
\addplot[cPink, dashdotted, mark=x, mark size=2.2pt, mark options={solid}] coordinates {(0,0.1245) (3,0.1245) (6,0.1245) (9,0.1245) (12,0.1245) (15,0.8887) (18,0.8887) (21,0.8887) (25,0.8887)};
\addlegendentry{HEVC}
\end{groupplot}
\node[anchor=north, yshift=-0.9cm] at (group c2r2.south) {\pgfplotslegendfromname{fig4legend}};
\end{tikzpicture}
\captionsetup{
    font=footnotesize,
    labelfont={bf,footnotesize}
}
\vspace{-5mm}
\caption{Reasoning-segmentation quality under matched communication budgets. Columns correspond to $K = 0.262$, $0.524$, and $1.311\times10^{6}$ channel uses per image; the upper and lower rows report gIoU and cIoU~\cite{lisa}, respectively. \texttt{HEVC} fails to decode below 13\,dB (gIoU $0.06$, cIoU $0.12$), so its curve drops below the plotted range. \texttt{FreshLatent} improves gIoU and cIoU over \texttt{Clean} by
20.79 and 20.87 points at 0\,dB and $K=0.262$M.}
\label{fig:matched}
\vspace{-5mm}
\end{figure*}
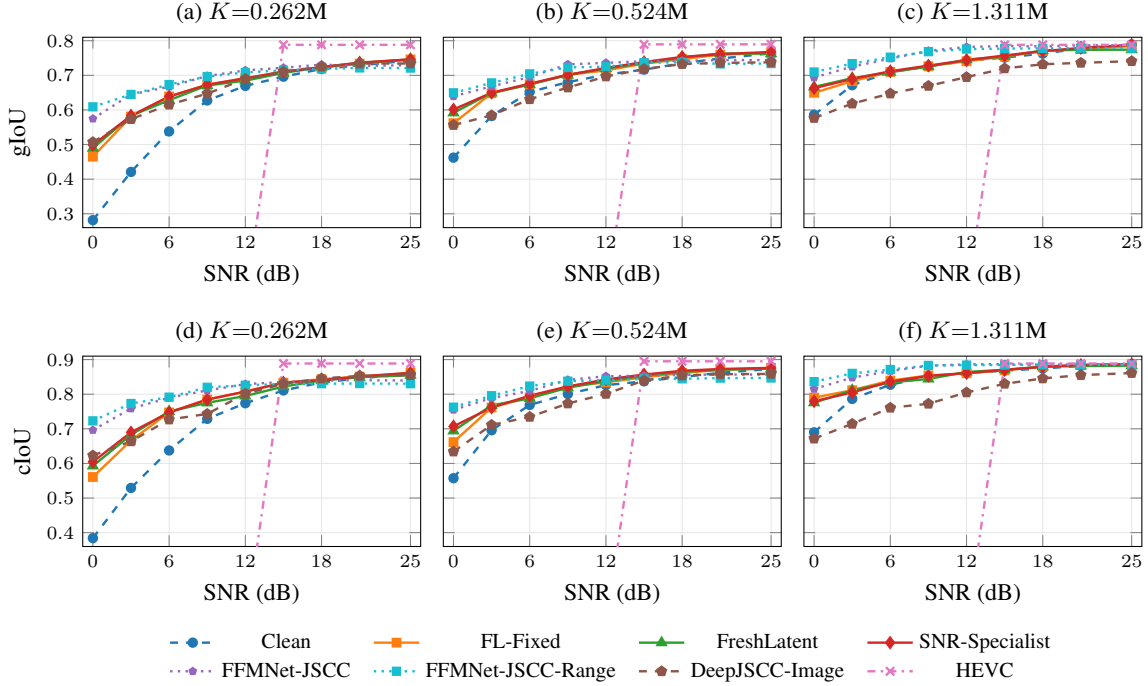

\underline{\textbf{(a) Model and datasets.}}
We use LISA-7B~\cite{lisa} with SAM ViT-H~\cite{sam}.
The latent interfaces are trained on ADE20K~\cite{ade20k} for
24 epochs using AdamW~\cite{adam} (learning rate $6\times10^{-5}$,
batch size 16), while the surrounding VLM remains frozen.
Downstream reasoning segmentation is evaluated using gIoU and cIoU
on the 200-image ReasonSeg validation set~\cite{lisa}.\\
\noindent
\underline{\textbf{(b) Split configuration.}}
Following~\cite{bhattacharjya2025avery}, we split the SAM image
encoder after its first transformer block, yielding
$F\in\mathbb{R}^{64\times64\times1280}$.
The reconstructed feature is passed to the remaining frozen LISA
pipeline.\\
\noindent
\underline{\textbf{(c) Compared methods.}}
We compare \texttt{FreshLatent} with the architecture-matched
\texttt{Clean} bottleneck, the fixed-13\,dB \texttt{FL-Fixed}
ablation, and the \texttt{SNR-Specialist} reference.
We further compare with \texttt{FFMNet-JSCC}, adapted from the
intermediate-feature JSCC architecture of~\cite{wang2021deep} to the
same SAM split and matched channel-use budgets, and
\texttt{FFMNet-JSCC-Range}, which uses the same FFMNet-JSCC
architecture but is trained over the same $\mathcal{U}[0,18]$\,dB
range as \texttt{FreshLatent}.
\texttt{FFMNet-JSCC-Range} provides the primary heavier-codec
comparison under matched range training, while the fixed-13\,dB
\texttt{FFMNet-JSCC} variant is retained to isolate the effect of
range training at unchanged codec capacity.
We also include \texttt{DeepJSCC-Image}~\cite{bourtsoulatze2019deepjscc}
and \texttt{HEVC}~\cite{hevc} image transmission. \\
\noindent
\underline{\textbf{(d) Communication and channel conditions.}}
We evaluate rates $r=\{0.05,0.10,0.25\}$, corresponding to
$K=\{0.262,0.524,1.311\}\times10^6$ real channel uses per image.
All methods are evaluated over AWGN SNRs
$\gamma\in\{0,3,6,9,12,15,18,21,25\}$\,dB at matched $K$. Training SNRs are $\mathcal{U}[0,18]$\,dB per batch for \texttt{FreshLatent} and
\texttt{FFMNet-JSCC-Range}, $\{1,4,7,13,19\}$\,dB for the \texttt{SNR-\allowbreak Specialist} bank, and 13\,dB for the remaining channel-aware methods. Additionally, at $r=0.10$, we evaluate the AWGN-trained \texttt{FreshLatent} under unseen Rayleigh fading without retraining.
 \\
\noindent
\underline{\textbf{(e) Embedded profiling.}}
We measure codec latency and energy on an NVIDIA Jetson AGX Xavier~\cite{nvidia_jetson_agx_xavier}
in \texttt{MODE\_10W}, a representative power-constrained embedded operating mode, with latency reported as the median of 20 runs.
\vspace{-5mm}
\subsection{Matched-Budget Perception Robustness}
\label{sec:accuracy_results}
\vspace{-1mm}
\underline{\textbf{(a) Benefit of channel-aware adaptation.}}
Fig.~\ref{fig:matched} shows that the gain over clean-trained
compression is largest when both channel quality and communication
budget are constrained.
At $K=0.262\times10^6$ and 0\,dB, \texttt{FreshLatent} improves
gIoU by 20.79 points and cIoU by 20.87 points over \texttt{Clean}.
As the communication budget increases, this gap narrows; at
$K=1.311\times10^6$, the corresponding 0-dB gIoU gain is 8.05 points.
Thus, channel-aware adaptation is most beneficial in the
communication-stressed regime.\\
\noindent
\underline{\textbf{(b) Range training versus specialization.}} For each latent rate, a single range-trained \texttt{FreshLatent} checkpoint is evaluated across all evaluated AWGN conditions. Across the resulting 27 $(K,\gamma)$ operating points, its mean absolute difference from \texttt{SNR-Specialist} is only 0.36 gIoU and 0.57 cIoU points, and it achieves higher gIoU at 8 points. Additionally, under unseen Rayleigh fading, the AWGN-trained \texttt{FreshLatent} shows partial cross-channel transfer, with average degradation of 7.28 gIoU and 8.92 cIoU points relative to AWGN over 0--25\,dB and gaps within 6.8 and 7.0 points, respectively, over 7--11\,dB.
\\
\noindent
\underline{\textbf{(c) Adaptation versus codec capacity.}}
\texttt{FFMNet-JSCC} provides additional robustness at several
poor-channel operating points, showing that greater codec capacity
can further improve perception quality.
At 0\,dB, \texttt{FL-Fixed}, trained at the same 13\,dB SNR as
\texttt{FFMNet-JSCC}, recovers approximately 56--63\% of its gIoU
improvement over \texttt{Clean} in the low- and medium-budget regimes.
Under matched range training, which controls for channel exposure, \texttt{FreshLatent} recovers
63.5--69.1\% of the gIoU improvement achieved by
\texttt{FFMNet-JSCC-Range} over \texttt{Clean} across all three
budgets at 0\,dB.
Within the heavier codec, range training shifts performance toward
poor-channel conditions: \texttt{FFMNet-JSCC-Range} exceeds its
fixed-13\,dB counterpart at 8 of the 9 points below 9\,dB, ties at
the remaining point, and is lower at all 18 points at or above 9\,dB,
with a maximum gain of 3.43 gIoU points at 0\,dB and the tightest
budget.
Sec.~\ref{sec:embedded_results} examines the interface cost required
for this additional robustness.\\

\noindent

\vspace{-1mm}

\vspace{-7mm}
\subsection{Mission-Conditioned Quality-Valid Region}
\label{sec:operating_space}

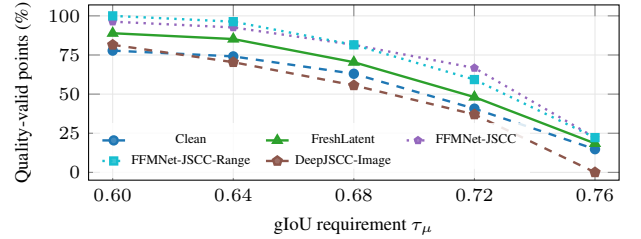
\begin{figure}[t]
\centering
\definecolor{cBlue}{RGB}{31,119,180}
\definecolor{cGreen}{RGB}{44,160,44}
\definecolor{cPurple}{RGB}{148,103,189}
\definecolor{cBrown}{RGB}{140,86,75}
\definecolor{cCyan}{RGB}{23,190,207}

\begin{tikzpicture}
\begin{axis}[
  width=\columnwidth,
  height=3.9cm,
  xlabel={gIoU requirement $\tau_\mu$},
  ylabel={Quality-valid points (\%)},
  xmin=0.592,
  xmax=0.768,
  ymin=-5,
  ymax=107,
  ytick={0,25,50,75,100},
  xtick={0.60,0.64,0.68,0.72,0.76},
  xticklabel style={
    /pgf/number format/fixed,
    /pgf/number format/zerofill,
    /pgf/number format/precision=2
  },
  grid=major,
  grid style={gray!20, line width=0.3pt},
  tick align=inside,
  tick label style={font=\scriptsize},
  label style={font=\scriptsize},
  every axis plot/.append style={line width=0.9pt},
  legend style={
    font=\tiny,
    draw=none,
    fill=white,
    fill opacity=0.85,
    text opacity=1,
    at={(0.42,0.01)},
    anchor=south,
    legend columns=3,
    /tikz/every even column/.append style={column sep=0.12cm},
    row sep=-1pt
  },
  legend image code/.code={
    \draw[mark repeat=2, mark phase=2, #1]
      plot coordinates {(0cm,0cm) (0.15cm,0cm) (0.3cm,0cm)};
  },
]

\addplot[
  cBlue, dashed,
  mark=*, mark size=1.5pt,
  mark options={solid}
]
coordinates {
  (0.60,77.8)
  (0.64,74.1)
  (0.68,63.0)
  (0.72,40.7)
  (0.76,14.8)
};
\addlegendentry{Clean}

\addplot[
  cGreen, solid,
  mark=triangle*, mark size=1.8pt,
  mark options={solid}
]
coordinates {
  (0.60,88.9)
  (0.64,85.2)
  (0.68,70.4)
  (0.72,48.1)
  (0.76,18.5)
};
\addlegendentry{FreshLatent}

\addplot[
  cPurple, dotted,
  mark=pentagon*, mark size=1.2pt,
  mark options={solid}
]
coordinates {
  (0.60,96.3)
  (0.64,92.6)
  (0.68,81.5)
  (0.72,66.7)
  (0.76,22.2)
};
\addlegendentry{FFMNet-JSCC}

\addplot[
  cCyan, dotted,
  mark=square*, mark size=1.3pt,
  mark options={solid}
]
coordinates {
  (0.60,100.0)
  (0.64,96.3)
  (0.68,81.5)
  (0.72,59.3)
  (0.76,22.2)
};
\addlegendentry{FFMNet-JSCC-Range}

\addplot[
  cBrown, dashed,
  mark=pentagon*, mark size=1.7pt,
  mark options={solid}
]
coordinates {
  (0.60,81.5)
  (0.64,70.4)
  (0.68,55.6)
  (0.72,37.0)
  (0.76,0.0)
};
\addlegendentry{DeepJSCC-Image}

\end{axis}
\end{tikzpicture}
\captionsetup{
    font=footnotesize,
    labelfont={bf,footnotesize}
}
\vspace{-4mm}
\caption{Fraction of the 27 evaluated $(K,\gamma)$ points satisfying
selected gIoU requirements $\tau_\mu$. \texttt{FL-Fixed} and
\texttt{SNR-Specialist} are omitted because they serve as an ablation
and condition-specific reference, respectively. \texttt{HEVC} is
omitted because decoding fails below 13\,dB, making its coverage
dominated by decoder availability rather than the quality requirement. \texttt{FreshLatent} increases quality-valid coverage over
\texttt{Clean} by 3.7--11.1 percentage points across all shown
requirements.}
\label{fig:valid_coverage}
\vspace{-7mm}
\end{figure}

\vspace{-2mm}
We next evaluate the quality component of
Eq.~\eqref{eq:task_validity}, instantiating $Q$ with gIoU.
Fig.~\ref{fig:valid_coverage} reports the fraction of the evaluated
$(K,\gamma)$ points satisfying five representative gIoU requirements:
0.60, 0.64, 0.68, 0.72, and 0.76.
Across these requirements, \texttt{FreshLatent} expands the
quality-valid region over \texttt{Clean} by 3.7--11.1 percentage
points.
At $\tau_\mu=0.72$ and 15\,dB, it meets the same requirement with
$K=0.524\times10^6$ rather than $1.311\times10^6$ channel uses,
reducing the minimum tested communication budget by 60\%.
Overall, these results show that \texttt{FreshLatent} broadens the
set of channel--communication conditions that satisfy
mission-conditioned perception requirements.

\vspace{-3mm}
\subsection{Embedded Interface Cost}
\label{sec:embedded_results}
\vspace{-1mm}
\underline{\textbf{Codec overhead.}}
\texttt{FreshLatent} uses \textbf{37--40$\times$ fewer encoder parameters}
($81.9$--$409.6\times10^{3}$ vs.\ $3.01$--$16.39\times10^{6}$),
\textbf{7.7--9.9$\times$ lower interface latency}
($3.79$--$14.95$\,ms vs.\ $29.21$--$147.93$\,ms), and
\textbf{8.8--10.0$\times$ lower interface energy}
($15.6$--$71.0$\,mJ vs.\ $136.7$--$713.1$\,mJ) than the heavier
\texttt{FFMNet-JSCC} codec across the three budgets.
The fixed- and range-trained variants share the same architecture and
therefore the same embedded interface cost.
Thus, the additional robustness of the heavier same-split codec comes
with substantially greater embedded cost, characterizing the
$\mathbf{C}^{M}(K)$ term in Eq.~\eqref{eq:task_validity}.\\


\noindent

\vspace{-3mm}
\noindent
\textit{\textbf{Takeaways.}}
The results reveal three effects.
First, at matched channel-use budgets, \texttt{Clean} achieves greater
quality-valid coverage than \texttt{DeepJSCC-Image} for all evaluated
requirements except $\tau_\mu=0.60$.
Second, under matched range training at 0\,dB, \texttt{FreshLatent}
recovers 63.5--69.1\% of the gIoU improvement achieved by the
substantially heavier \texttt{FFMNet-JSCC-Range} codec across the
three communication budgets.
Finally, these robustness gains translate into a broader quality-valid
channel--communication region, enabling \texttt{FreshLatent} to satisfy
mission-conditioned perception requirements across more constrained
operating conditions.
\vspace{-4mm}
\section{Conclusion and Future Work}
\vspace{-3mm}
We presented \texttt{FreshLatent}, a lightweight channel-aware
adaptation of the communication interface in split embodied VLMs.
Under matched range training, \texttt{FreshLatent} recovers a
substantial fraction of the poor-channel robustness gain of a heavier
feature-JSCC codec while expanding the quality-valid
channel--communication region over clean split inference.
On an NVIDIA Jetson AGX Xavier, \texttt{FreshLatent} uses
37--40$\times$ fewer encoder parameters, 7.7--9.9$\times$ lower
interface latency, and 8.8--10.0$\times$ lower interface energy than
the heavier same-split alternative.
These results show that adapting a lightweight split interface to
wireless corruption can provide substantial robustness gains without
a much heavier codec.
Future work will move beyond static interface adaptation toward
runtime communication--computation control under changing operating
conditions~\cite{bhattacharjya2026access} and selective fallback when
communication-dependent reasoning becomes unreliable~\cite{bhattacharjya2026memoguard},
together with operator-level evaluation.

\bibliographystyle{IEEEbib}
\bibliography{references}

\end{document}